\documentclass[reprint,aps,prl,superscriptaddress]{revtex4-2}

\usepackage{amsmath, amssymb, times, mathrsfs, hyperref, array, bbm}
\usepackage{graphicx}
\usepackage[usenames,dvipsnames]{xcolor}
\usepackage{braket}
\usepackage{xcolor}
\usepackage{tcolorbox}
\usepackage{listings}
\hypersetup{colorlinks=false}
\usepackage{hyperref}

\usepackage{ulem}
\usepackage{lipsum}
\usepackage{listings}

\usepackage{mdframed}
\usepackage{balance}

\begin{document}
\title{Reply to the Comment by Tikhonov and Khrapai on ``Long-range crossed Andreev reflection in a topological insulator
nanowire proximitized by a superconductor”}

\author{Junya Feng}
\affiliation{Physics Institute II, University of Cologne, Z\"ulpicher Str. 77, 50937 K\"oln, Germany}
\author{Henry F. Legg}
\affiliation{SUPA, School of Physics and Astronomy, University of St Andrews, North Haugh, St Andrews, KY16 9SS, United Kingdom}
\affiliation{Department of Physics, University of Basel, Klingelbergstrasse 82, CH-4056 Basel, Switzerland}
\author{Mahasweta Bagchi}
\affiliation{Physics Institute II, University of Cologne, Z\"ulpicher Str. 77, 50937 K\"oln, Germany}
\author{Daniel Loss}
\affiliation{Department of Physics, University of Basel, Klingelbergstrasse 82, CH-4056 Basel, Switzerland}
\author{Jelena Klinovaja}
\affiliation{Department of Physics, University of Basel, Klingelbergstrasse 82, CH-4056 Basel, Switzerland}
\author{Yoichi Ando}
\affiliation{Physics Institute II, University of Cologne, Z\"ulpicher Str. 77, 50937 K\"oln, Germany}

\begin{abstract}
The comment (arXiv:2505.23490) fails to identify any scientific errors and its central arguments actually support the main conclusions of our publication [Nat. Phys. {\bf 21}, 708 (2025)]. Firstly, the whole argument of the comment to try to explain our data explicitly relies on the existence of a large crossed Andreev reflection (CAR) effect. 
The presence of a sizable CAR transmission probability over a surprisingly long distance is the first conclusion of our publication. Secondly, the comment discusses the complex interplay of CAR and ECT, especially in the presence of local effects. This complex interplay is precisely the second conclusion of our publication. In essence, the comment amounts to merely pointing out that there is a broader sense in the notion of ``dominant CAR'' when nonlinear effects become relevant.
\end{abstract}

\maketitle
\noindent In our recent article [Feng et al.~Nat.~Phys. {\bf 21}, 708–715 (2025)]~\cite{Feng2025} we reported the presence of a long-range crossed-Andreev reflection (CAR) contribution to the nonlocal conductance in a proximitised topological insulator (TI) nanowire. Our work also highlighted the intricate interplay of CAR and elastic co-tunnelling (ECT) transmission in mesoscopic superconducting systems. In particular, the conclusion of Ref.~\citenum{Feng2025} states that nonlocal conductance ``reflects a complex interplay of local and global properties".

Recently, a comment by Tikhonov and Khrapai~\cite{Tikhonov2025} discussed the results of Ref.~\citenum{Feng2025}. Confusingly, this comment actually explicitly supports our two central results. 
%Firstly, the authors argue that, in similar devices~\cite{Denisov2021,Denisov2022}, CAR and ECT transmission probabilities are nearly equal on average ($T_{12}^{he} \approx T_{12}^{ee}$) and that our data ``closely resembles'' the data from those similar devices. The presence of a CAR transmission probability that competes and can dominate over ECT is the core result of Ref.~\citenum{Feng2025}.
%Firstly, the whole argument of Tikhonov and Khrapai to try to explain the data reported in Ref.~\citenum{Feng2025} relies on the existence of a finite $T_{12}^{he}$, namely, the existence of  CAR. The presence of a finite CAR transmission probability over a surprisingly long distance is the first conclusion of Ref.~\citenum{Feng2025}.
Firstly, the whole argument of Tikhonov and Khrapai to try to explain the data reported in Ref.~\citenum{Feng2025} explicitly utilizes a finite $T^{he}_{12}$, namely, it requires the existence of CAR. Moreover, in their theory this CAR contribution is similar in magnitude to the ECT contribution $T_{12}^{ee}$, such that $T_{12}^{he} \approx T_{12}^{ee}$, which implies a large CAR contribution. The presence of a sizable CAR transmission probability over a surprisingly long distance is the first conclusion of Ref.~\citenum{Feng2025}.
Secondly, Tikohnov and Khrapai discuss the complex interplay of CAR and ECT, especially in the presence of local effects. As noted above, this complex interplay is precisely the second conclusion of Ref.~\citenum{Feng2025}. 

In their comment, Tikhonov and Khrapai primarily discuss Fig.~3(e-f) of our work. In particular our statement that: ``The competition between ECT and CAR can be manipulated by applying a finite bias not only to the left, but also to the right''. In fact, the comment concludes that ``the core idea of experiment [1] about the bias combination controllability of the CAR and ECT contributions in nonlocal conductance is misleading.'' We disagree with this assertion.

Based on a simplified model with bias-independent transmission, the comment starts its discussion by saying that we incorrectly cited Ref.~\citenum{Bordin2023} as a demonstration of CAR and ECT manipulation by bias combination, because Ref.~\citenum{Bordin2023} primarily discusses total current, rather than differential conductance. However, this neglects the fact that the figures in Ref.~\citenum{Bordin2023}  also clearly show a dependence of the magnitude of current on bias voltage combination, i.e., the differential conductance. Furthermore, the comment ignores that all of our figures show a nonlocal conductance with a considerable bias dependence that is clearly incompatible with their bias-independent model (as the rest of their comment also acknowledges).
%Such a criticism based on an unrealistic model is not scientifically valid.

Next, the authors add bias-dependent transmission probabilities to their model and perform a Taylor expansion of the nonlocal conductance with respect to the left and right bias voltages. From this they argue that certain symmetry-allowed nonlinear terms explain the data in Fig.~3(e-f), in particular ``self-gating'' terms. 
%Although such terms are certainly allowed and so will be present, we note that the authors do provide a model of transmission probabilities that would produce similar data. Most importantly, this expansion, does not contradict our statement that the competition between CAR and ECT can be manipulated by bias. 
It is important to note here that this expansion does not contradict our statement that the competition between CAR and ECT can be manipulated by bias.
In fact, one does not need to resort to such a Taylor expansion. The nonlocal conductance can -- in principle -- be directly calculated in terms of the voltage and energy dependent CAR ($T_{12}^{he}$) and ECT ($T_{12}^{ee}$) transmission probabilities. The comment itself provides the equation to do this [Eq.~(5) of the comment]
\begin{equation}
\frac{G_{12}}{G_{0}}=\frac{1}{e}\frac{\partial}{\partial V_{2}}\int_{0}^{eV_{2}}[T_{12}^{he}(E,V_{1},V_{2})-T_{12}^{ee}(E,V_{1},V_{2})]dE,\label{nonlocalconductance}
\end{equation}
with $\frac{G_{12}}{G_{0}}$ the (normalised) nonlocal conductance, $V_1$ and $V_2$ the bias voltages applied on either side, and $E$ the energy. The form in Eq.~\eqref{nonlocalconductance} allows for a direct discussion of the competition between the CAR and ECT contributions (averaged over energy), rather than obscuring them in a Taylor expansion and providing each term with a different name. Tikhonov and Khrapai's comment amounts to pointing out that there is a broader sense in the notion of ``dominant CAR'' in Eq.~\eqref{nonlocalconductance} when nonlinear terms in their expansion become relevant.

Finally, the comment describes our observation of bias symmetry as a ``fine-tuned'' effect. This framing entirely ignores the underlying message of our manuscript. What the comment calls ``fine-tuning'' is precisely the dependence of nonlocal conductance on gate and bias configurations that our paper discusses. We were clear throughout our work that there is a complex interplay between CAR and ECT contributions that is highly dependent on gate and bias voltages. This strong dependence on device configuration is, in fact, the central concluding message of our paper. As such, the comment's analysis, rather than refuting our results, simply reaffirms our conclusion that (i) there is a sizable long-range CAR component and (ii) nonlocal processes are highly sensitive functions of the system parameters.

\vspace{-14pt}

\end{document}